\documentclass[prl,reprint,amsmath,amssymb,showpacs,twocolumn]{revtex4-1}

\usepackage[dvipdfmx]{graphicx}
\usepackage{bm}
\usepackage{mathrsfs}
\usepackage{color}

\DeclareMathOperator{\diag}{diag}

\begin{document}
\title{Topological superconductivity in Dirac semimetals}
\author{Shingo Kobayashi}
\author{Masatoshi Sato}
\affiliation{Department of Applied Physics, Nagoya University, 
Nagoya 464-8603, Japan}

\date{\today}

\begin{abstract}
Dirac semimetals host bulk band-touching Dirac points and
a surface Fermi loop.  
We develop a theory of superconducting Dirac semimetals. 
Establishing a relation between the Dirac points and the 
 surface Fermi loop, we clarify how the
 nontrivial topology of Dirac semimetals affects their superconducting
 state. 
We note that the unique orbital texture of Dirac points and 
a structural phase transition of the crystal favor 
symmetry-protected topological superconductivity with a quartet of surface Majorana
 fermions.
We suggest possible application of our theory to 
recently discovered superconducting states in
 Cd$_3$As$_2$.

\end{abstract}
\pacs{}
\maketitle
Dirac semimetals are three-dimensional (3D) materials that possess
gapless (Dirac) points in the bulk Brillouin zone (BZ), whose low-energy
excitations are effectively described as Dirac fermions. With
time-reversal symmetry (TRS) and inversion symmetry (IS) preserved, a pair of
Dirac points is formed at the crossing of two doubly degenerate
bands on a high-symmetry axis. 
They are protected by discrete rotation
($C_n$) symmetry~\cite{Young:2012,Wang:2012,Wang:2013,BJYang:2014},
which prohibits band mixing to open a gap. 
Furthermore, Dirac semimetals may host a surface Fermi loop
(FL)~\cite{BJYang:2014,Wang:2013,Yi:2014,Neupane:2015}. This contrasts sharply with a surface
Fermi arc in Weyl semimetals~\cite{Wan:2011} because its
topological origin is different. Several Dirac semimetals, including
Na$_3$Bi~\cite{Liu:2014a,Xu:2013,Xu:2015,Kushwaha:2015} and
Cd$_3$As$_2$~\cite{Neupane:2014,Borisenko:2014,Yi:2014,Liu:2014b,Jeon:2014,He:2014,Neupane:2015,Liang:2015},
have been demonstrated experimentally and predicted
theoretically~\cite{Steinberg:2014,Gibson:2014,Du:2014,Seibel:2015,Narayan:2014}.  

Superconducting phase transitions were reported recently in
Cd$_3$As$_2$~\cite{Aggrawal:2014,Wang:2015,He:2015} and
Au$_2$Pb~\cite{Schoop:2014}, both of which support Dirac points
protected by $C_4$ symmetry. 
Bulk Cd$_3$As$_2$ exhibits superconductivity 
under high pressure ($\sim8.5$ GPa)~\cite{He:2015}
accompanied by a structural phase transition of the
crystal~\cite{Zhang:2014}. 
In addition, point contact measurements of
Cd$_3$As$_2$ reportedly
induce superconductivity around the point contact region, where
the tunneling conductance shows a
zero-bias conductance peak~\cite{Aggrawal:2014,Wang:2015}. 
Au$_2$Pb also exhibits a superconducting phase transition 
after a structural phase transition~\cite{Schoop:2014}.

In this letter we address the effect of the nontrivial
topology, i.e., the Dirac points and FL, on the superconducting
properties. Topological materials are a promising
platform to realize topological superconductors (TSCs) owing to the
nontrivial topology of the wave function in normal states~\cite{Sato:2003,Fu:2008,Fu:2010,Qi:2010,Yamakage:2012,Lu:2014}. 
For instance, surface Dirac fermions may realize a TSC even for
an $s$-wave pairing state~\cite{Sato:2003,Fu:2008}.
Also, the Fermi
surface topology, which is the simplest topological structure in the
normal state,  directly affects the topological
superconductivity of odd-parity superconductors~\cite{Sato:2010,Fu:2010}. 
For the carrier-doped  topological insulator,
topological superconductivity has
been anticipated for the surface~\cite{Fu:2008} or the bulk~\cite{Fu:2010}. 

Here we present a general framework for studying
superconductivity in Dirac semimetals. 
The key ingredients are symmetry-protected topological numbers in
crystalline insulators and superconductors~\cite{Fu:2011,Chiu:2013,Morimoto:2013,Shiozaki:2014,Kobayashi:2014,Ueno:2013,FZhang:2013,Fang:2012b,Alexandradinata:2014,Slager:2013,SAYang:2014,Koshino:2014}. 
In particular, we examine the $C_4$ topological invariant and the mirror
Chern number, which ensure the existence of Dirac points and FLs, respectively, in Dirac
semimetals. 
First, we show that these two 
topological numbers are intrinsically related to each other,
establishing a relation between Dirac points and surface FLs.
Then, 
we elucidate how the nontrivial topology of Dirac semimetals affects
their superconducting state.
We find that for a class of pairing symmetries, 
Dirac points and FLs in the normal state are
inherited as bulk point nodes and surface Majorana fermions (MFs), respectively, in the
superconducting state.

By carefully examining the low-energy
effective Hamiltonian, we also reveal that doped Dirac semimetals
favor an equal-spin odd-parity pairing rather than a
conventional $s$-wave one. 
The former pairing exhibits a distinct quartet of surface MFs
stemming from the FL,
though point nodes exist when the system retains
$C_4$ rotation symmetry.
If the $C_4$ symmetry is reduced to
$C_2$ by a
structural phase transition, the nodes disappear, and a full-gapped symmetry-protected TSC
is realized.
The FL-induced MFs are clearly distinguished from those in other
TSCs~\cite{Tanaka:2012,Qi:2011} including superfluid
$^3$He-A~\cite{Kopnin:1991,Volovik:2013} and Weyl
superconductors~\cite{Meng:2012,GYCho:2012,Hosur:2014,Lu:2014,Shivamoggi:2013}.  
We finally suggest possible application of our theory to 
recently discovered superconducting states in
 Cd$_3$As$_2$ and Au$_2$Pb.       

\begin{figure}[tbp]
\centering
 \includegraphics[width=8cm]{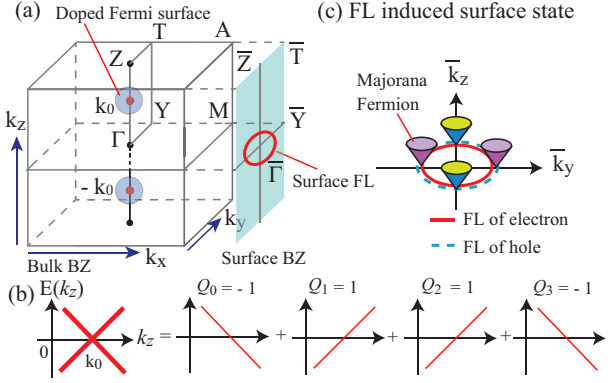}
 \caption{(Color online) (a) Schematic picture of the Dirac points at $(0,0,\pm k_0)$ and
 the surface Fermi FL in the bulk BZ and the surface BZ,
 respectively. (b) On the $k_z$-axis, a bulk Dirac point
 with $Q=-1$ in the left-hand side is
 decomposed into four chiral modes with different $\alpha_p$ in the
 right-hand side. The bold line is doubly degenerate.
 (c) Schematic illustration of the surface Majorana quartet in the superconducting Dirac semimetal.}\label{fig:TDS}
\end{figure}

{\it Stability of Dirac points and surface Fermi loop}---
First, we provide a general argument on the topology of Dirac points.
Our theory assumes TRS, IS, and uniaxial rotation symmetry, which are the most
common symmetries for Dirac semimetals.
In the presence of TRS and IS, Kramer's degeneracy exists at
arbitrary $\bm{k}$ in the BZ,
ensuring fourfold degeneracy when the conduction and valence bands
are in contact. Such an accidental band crossing is generally not stable owing
to band repulsion. However, if the band-touching point is on the high-symmetry 
axis, a $C_n$ symmetry can retain the band
crossing as a Dirac point~\cite{Young:2012,Wang:2013,Wang:2012,BJYang:2014}. Below we clarify the relevant topological structures.

We focus on the $C_4$ symmetric Hamiltonian, 
$
C_4 H(\bm{k}) C_4^{-1} = H(R_4 \bm{k}),
$
with $R_4 \bm{k} = (k_y,-k_x,k_z)$. 
Here, without losing generality, we have chosen the rotation axis as
the $k_z$ axis. 
The commutation relations between the TRS operator $T$, IS operator $P$, and $C_4$ are summarized as
$[T,P]=[T,C_4]=[P,C_4]=0$. 
As illustrated in Fig.~\ref{fig:TDS}(a),
for a $C_4$ symmetric tetragonal crystal, there are two
$C_4$ symmetry
lines, $\Gamma Z=(0,0,k_z)$ and $M A =(\pi,\pi,k_z)$, with 
$k_z \in [-\pi,\pi]$. 
On these $C_4$ symmetry lines, the Hamiltonian commutes
with $C_4$; thus, 
any energy band along the high-symmetry 
lines has a definite eigenvalue of the $C_4$
operator, $\alpha_{p}= \exp \left[ \frac{i
\pi}{2} \left(p+\frac{1}{2} \right) \right]$ ($p=0,1,2,3$).

The existence of Dirac points is ensured by the topological invariant
defined below.
The system is generally
gapped at the $C_4$ symmetry points $k=\Gamma, Z, M, A$, so one can
count the number of bands below the Fermi level at these points.
Denoting the number of such bands with the $C_4$
eigenvalue $\alpha_p$ at $k$ as $N_p(k)$, we can
introduce the topological number, 
$\mathcal{Q}_p = N_p(\Gamma) - N_p(Z),$
for Dirac points on the $\Gamma Z$ line,
which we call the $C_4$ topological invariant.
The Kramer's degeneracy due to $PT$ symmetry requires $\mathcal{Q}_{p} =
\mathcal{Q}_{3-p}$ $(p=0,1)$. 
Moreover, 
the sum rule $\sum_{p=0}^{3}
\mathcal{Q}_p =0$ holds for Dirac semimetals.
Indeed, for a Dirac semimetal to become an insulator by a small
$C_4$ breaking perturbation, the total number of bands below the Fermi
level should be the same at all the symmetry points $k$, which leads
to the sum rule.  
From these two relations, we have 
\begin{align}
\mathcal{Q}_0=-
\mathcal{Q}_1=- \mathcal{Q}_2= \mathcal{Q}_3\equiv \mathcal{Q}.
\label{eq:constraint}
 \end{align}
 If $\mathcal{Q}\neq 0$, any band with $\alpha_p$ has
$\mathcal{Q}$ gapless points on the $\Gamma Z$ lines, 
corresponding to the difference $N_p(\Gamma)-N_p(Z)$,   
which eventually form
$\mathcal{Q}$ Dirac points.
In Fig.~\ref{fig:TDS}(b), we illustrate a Dirac point with $Q=-1$.
Because band mixing between different $C_4$ eigensectors is
prohibited, the resultant Dirac points are stable as long as 
$C_4$ symmetry is maintained. 
Similarly, we can also introduce the $C_4$ invariant for Dirac
points on the $MA$ line. 

Because of IS and the $C_2$ subgroup for $C_4$ symmetry,  
the system also has
mirror reflection symmetry, $M_{xy} H(k_x,k_y,k_z) M_{xy}^{-1}
=H(k_x,k_y,-k_z)$, with $M_{xy}= C_4^2 P$.
In the mirror-invariant planes ($k_z = 0$ or $\pi$), 
the Hamiltonian is block-diagonal in the basis of the eigenstates of 
$M_{xy}$; thus,
the mirror Chern number $\nu_{\lambda}(k_z)$ with $k_z=0, \pi$ is
defined as
 $\nu_{\lambda} (k_z) :  = 1/2\pi \int_{\rm BZ} dk_x dk_y \mathcal{F}^{\lambda} (\bm{k})$, with $ \mathcal{A}_a^{\lambda}(\bm{k}) := \sum_{E_n <0} i \langle u_n^{\lambda} (\bm{k}) | \partial_{k_a} u_n^{\lambda} (\bm{k}) \rangle $~\cite{Teo:2008,Heish:2012}, where $|u_n^{\lambda} (\bm{k}) \rangle $ is an eigenstate of $H(\bm{k})$ in the mirror sector with the eigenvalue $\lambda=\pm i$ of $M_{xy}$ and $\mathcal{F}^{\lambda}$ is the field strength of $\mathcal{A}_a^{\lambda}$.
Generalizing the relation between a band inversion and the Chern number in terms of eigenvalues of crystal symmetry~\cite{Fu:2007,Sato:2009,Fang:2012,Fang:2012b,Benalcazar:2014},
we obtain the following relation for the mirror Chern number:
\begin{align}
 e^{\frac{i \pi}{2}\nu_{\lambda}(0)} = \prod_p 
\alpha_{p}^{[N_{(p, \lambda)}(\Gamma)+N_{(p,\lambda)}(M)]}\prod_q
\xi_{q}^{-{\cal N}_{(q, \lambda)}(Y)}, 
\label{eq:nuM}
\end{align}
where $\xi_q=\exp\left[i\pi\left(q+\frac{1}{2}\right)\right]$ $(q=0,1)$
is the eigenvalue of $C_2$, 
$N_{(p,\lambda)}(k)$ is the
number of occupied bands at $k$ with a set of the $C_4$ and $M_{xy}$
eigenvalues $(\alpha_p,
\lambda)$, and ${\cal N}_{(q,\lambda)}(Y)$ are those at $Y$ with a set
of $C_2$ and $M_{xy}$ eigenvalues $(\xi_q, \lambda)$.
[$Y$ is the $C_2$ symmetry point in
Fig.~\ref{fig:TDS}(a)].
Note that occupied bands at the $C_4$ ($C_2$) symmetry points have a
definite set of eigenvalues for $C_4$ ($C_2$) and $M_{xy}$ because $[C_4,
M_{xy}]=[C_2, M_{xy}]=0$. 
We can also obtain a similar relation for $\nu_{\lambda}(\pi)$ by
replacing $\Gamma$, $M$, and $Y$ with $Z$, $A$, and $T$, respectively, in
Fig.~\ref{fig:TDS}(a).

To see the close relationship between the $C_4$ invariant and the
mirror Chern number, consider a process in
which a pair of stable Dirac points is created at $\Gamma$
\cite{BJYang:2014}. 
Band inversion at $\Gamma$ occurs in this process, so  
a Kramer's pair of occupied bands, which have a set of eigenvalues
$(\alpha_{p'},\lambda)$ and $(\alpha_{3-p'},-\lambda)\equiv(\alpha_{p'}^*,
\lambda^*)$,
go above the Fermi level, and a Kramer's pair of empty bands with
the eigenvalues $(\alpha_{p''}, \lambda)$ and $(\alpha_{3-p''},
-\lambda)$ go below it at $\Gamma$. 
As a result, $N_{(p,\lambda)}(\Gamma)$ changes by $\Delta
N_{(p',\lambda)}(\Gamma)=\Delta N_{(3-p', -\lambda)}(\Gamma)=-1$, $\Delta
N_{(p'', \lambda)}(\Gamma)=\Delta N_{(3-p'', -\lambda)}(\Gamma)=1$.
To have a stable Dirac point, $\mathcal{Q}_p$ should change at the same
time, so $p'\neq p''$. 
Then, from Eq. (\ref{eq:nuM}), we find that this process induces
a simultaneous change in the  mirror Chern number $\Delta \nu_{\lambda}(0)$:
 \begin{align}
  \Delta \nu_{\lambda}(0) &= \sum_p\left( p + \frac{1}{2} \right)
  \Delta N_{(p,\lambda)} (\Gamma) \mod 4,
\nonumber\\ 
&=(p''-p')\neq 0 \mod 4.
 \end{align}
Therefore, the creation of stable Dirac points is always accompanied by
a net change in the mirror Chern number. 
From the bulk-boundary correspondence, the resultant mirror Chern number ensures the existence of surface
helical Dirac fermions, whose Fermi surfaces form FLs.

Although the surface FL accompanies bulk Dirac points, it can be stable
even when $C_4$ symmetry is lost, so the Dirac points have gaps.
Indeed, unless the $C_2$ subgroup is broken, the system maintains mirror
reflection symmetry, which is sufficient to stabilize the surface FL. 
Therefore, the structural phase transition that breaks $C_4$ to $C_2$
retains the FL.

{\it Topology of superconducting Dirac semimetals}---  
With a finite carrier density, Dirac semimetals have disconnected bulk
Fermi surfaces, each of which surrounds one of the band-touching Dirac
points. See Fig.~\ref{fig:TDS}(a).  
Now consider a superconducting state in Dirac semimetals.
The system is described by the Bogoliubov--de Gennes (BdG) Hamiltonian, 
 \begin{align}
H_{\rm BdG} (\bm{k})= 
\begin{pmatrix} 
H(\bm{k})-\mu  &
\Delta(\bm{k})\\ \Delta^{\dagger} (\bm{k}) & 
-H^{\ast} (-\bm{k})+\mu
\end{pmatrix} ,
\label{eq:BdG2}
 \end{align}  
where $H(\bm{k})$ is the Hamiltonian for Dirac semimetals discussed above, $\mu$
is the chemical potential corresponding to the finite carrier density, and
$\Delta({\bm k})$ is the gap function.
The BdG Hamiltonian supports particle--hole symmetry, 
$CH_{\rm BdG}(\bm{k}) C^{-1} = - H_{\rm BdG}(-\bm{k})$,
$C=\tau_x K$,
with the Pauli matrix $\tau_x$ in the Nambu space and the conjugation
operator $K$.
Moreover, it may retain the symmetries of Dirac semimetals, 
depending on the symmetry property of the gap function.
In particular, for a gap function with  $C_4 \Delta (\bm{k}) C_4^t = e^{-
\frac{ i \pi r }{2}} \Delta (R_4 \bm{k})$ ($r=0,\cdots, 3$),
$H_{\rm BdG}({\bm k})$ keeps $C_4$ symmetry, $\tilde{C}_4
H_{\rm BdG}({\bm k})\tilde{C}_4^{-1}=H_{\rm BdG}(R_4{\bm k})$, with 
$\tilde{C}_4 = \diag [ C_4,  e^{\frac{ i \pi r }{2}} C_4^{\ast}
]$. 
In addition, for a mirror-even or mirror-odd gap function that satisfies
$M_{xy}\Delta(\bm{k}) M_{xy}^t= \eta_{\rm M} \Delta(k_x, k_y, -k_z)$
with $\eta_{\rm M}=\pm 1$,
the system retains mirror reflection symmetry, $\tilde{M}_{xy}H_{\rm BdG}({\bm
k})\tilde{M}^{-1}_{xy}=H_{\rm BdG}(k_x, k_y, -k_z)$, with
$\tilde{M}_{xy}=\diag[M_{xy}, \eta_{\rm M} M^*_{xy}]$.
Correspondingly, we can introduce the $C_4$-invariant
$\tilde{Q}_p$ and the
mirror Chern numbers $\tilde{\nu}_{\lambda}$ for $H_{\rm BdG}({\bm k})$~\cite{Ueno:2013,FZhang:2013},
in a manner similar to that used for those of $H({\bm k})$. 
The topological numbers $\tilde{Q}_p$ and $\tilde{\nu}_{\lambda}$ are
responsible for the existence of bulk point nodes on the $\Gamma Z$ line
and surface MFs in the superconducting state, respectively.
  
To evaluate these topological numbers,
we employ the weak pairing assumption~\cite{Fu:2010,Sato:2010,Qi:2010}, i.e.,
that the superconducting gap is much smaller than the Fermi energy. 
The gap function is reasonably negligible away from the Fermi
surface, in which we can take $\Delta (\bm{k}) \to 0$, leading to
$H_{\rm BdG}(\bm{k}) \to
\diag[H(\bm{k})-\mu,-H^{\ast}(-\bm{k})+\mu]$.
Therefore, at the symmetry points $k=\Gamma, Z, M, A$, we can relate
the negative energy states of $H_{\rm BdG}(k)$ to those
of $H(k)$.
By taking into account the contribution from holes as well as electrons, 
the number
$\tilde{N}_p$ of negative energy states with the $\tilde{C}_4$
eigenvalue $\alpha_p$ is evaluated as
$   \tilde{N}_{p} (k)= N_{p} (k) +[N- N_{p_h} (k)],$
where $N$ is the total number of bands in $H({\bm k})$, $p_h=3 -
p + r \mod 4$, and the first (second) term on the
right-hand side comes
from the electron (hole) contribution.
From this equation, the $C_4$ invariant in the superconducting state
is obtained as 
\begin{align}
 \tilde{\mathcal{Q}}_p = \begin{cases} 0 & r=0 \text{ or } p=p_h, \\ 2 \mathcal{Q}_p &{\rm otherwise}. \end{cases} \label{eq:rel-Q}
\end{align} 
Similarly, the mirror Chern number in the superconducting state is
calculated as the sum of the electron and hole mirror Chern numbers,
$
 \tilde{\nu}_{\lambda} = \nu_{\lambda} + \nu_{\lambda_h}, 
$
with $\lambda_h=-\eta_{\rm M}\lambda$.
As TRS in Dirac semimetals implies  $\nu_{-\lambda} = -
\nu_{\lambda}$, 
we have 
\begin{align}
 \tilde{\nu}_{\lambda} = \begin{cases} 0 & \eta_{\rm M}=1, 
\\ 2 \nu_{\lambda} &\eta_{\rm M}=-1. \end{cases} 
\label{eq:rel-nu}
\end{align} 

Relations (\ref{eq:rel-Q}) and (\ref{eq:rel-nu}) have important
physical consequences. (i) In the presence of $C_4$ symmetry, any
superconducting Dirac semimetal with a nontrivial $r$ ($r=1,2,3$)
hosts point nodes as a remnant of Dirac points.
Indeed, Eqs. (\ref{eq:constraint}) and (\ref{eq:rel-Q}) imply that at
least a couple of $\tilde{Q}_p$ are nonzero in this case.
To open a point node gap, we need to break the $C_4$ rotation symmetry.
(ii) If the gap function is mirror-odd, the mirror Chern number of
the superconducting Dirac semimetal is nonzero, resulting in double
MFs. 
In Fig.~\ref{fig:TDS}(c), we illustrate how the double MFs are created.
In general, the gap function mixes the surface FL
of electrons with that of holes so as to open a gap for the FL.   
In the mirror-odd case, however,
mixing is prohibited on the mirror-invariant line in the surface BZ, so a
pair of gapless points remains for each FL, forming double MFs. 
 
In addition to the double MFs on the mirror-invariant line, we also find that each FL in the mirror-odd superconductor creates
another pair of MFs {\it on the $k_z$ axis in the surface BZ}. 
By combining with $C$ and $\tilde{M}_{xy}$, the BdG Hamiltonian 
for the surface FL on the $k_z$ axis, which we denote
$H_{\rm BdG}^{\rm FL}(k_z)$, has antiunitary antisymmetry,
$C_M H_{\rm BdG}(k_z) C_M^{-1}=-H_{\rm BdG}^{FL}(k_z)$, with
$C_M=i\tilde{M}_{xy}C=i\tilde{M}_{xy}\tau_x K$. 
Because $C_M^2=1$ in the mirror-odd superconductor, 
$H_{\rm BdG}^{FL}(k_z)i\tilde{M}_{xy}\tau_x$ 
is found to be real antisymmetric; thus, by using the Pfaffian, 
we can introduce the zero-dimensional topological invariant 
$\chi(k_z)={\rm sgn}\{{\rm Pf}[H_{\rm
BdG}^{FL}(k_z)i\tilde{M}_{xy}\tau_x]\}$.
In the weak pairing case, $\chi(k_z)$ is evaluated as
$\chi(k_z)={\rm sgn}\{{\rm det}[H^{FL}(k_z)-\mu]\}$, where $H^{\rm
FL}(k_z)-\mu$ is the Hamiltonian of the surface FL on the $k_z$ axis \cite{SM}. 
Therefore, $\chi(k_z)$ has different signs inside and
outside the FL, which implies that $H_{\rm BdG}^{FL}(k_z)$ should have
zero-energy states near the points of intersection between the FL and the
$k_z$ axis.   
These zero-energy states form a pair of MFs on the $k_z$ axis. 
Consequently, we can conclude that each FL has a quartet of
MFs, as shown in
Fig.~\ref{fig:TDS}(c). 
The quartet of MFs can stay gapless even when $C_4$ symmetry
is broken, as long as mirror symmetry is preserved.

{\it Low-energy analysis and application to Cd$_3$As$_2$.}---For definiteness, we study the low-energy effective Hamiltonian, which
describes a class of Dirac semimetals including
Cd$_3$As$_2$ and Au$_2$Pb.
Because bands in Dirac semimetals are doubly degenerate owing to $PT$ symmetry,
band-touching Dirac points are minimally described by a $4\times
4$ matrix Hamiltonian. 
Thus, in the minimal setup, we need orbital degrees of freedom in addition to 
spin degrees of freedom, 
which are given by the
Pauli matrices $\sigma_{\mu}$ and $s_{\mu}$ in the orbital $(1,2)$ and spin $(\uparrow,\downarrow)$ spaces, respectively. 
The form of the $4\times 4$ Hamiltonian is uniquely
determined by symmetry~\cite{Wang:2013,BJYang:2014}.
In particular, for $P=\pm \sigma_z$, the low-energy
lattice Hamiltonian is given by \cite{BJYang:2014}
\begin{align}
 H  (\bm{k}) 
=&   \{ M - t_{xy}(\cos k_x +\cos k_y ) 
-  t_z \cos k_z \}  \sigma_z s_0 \notag \\
&+ (\eta \sin k_x) \sigma_x s_z  -  (\eta \sin k_y) \sigma_y s_0  \notag \\
& + (\beta + \gamma)\sin k_z(\cos k_y -\cos k_x) \sigma_x s_x \notag \\
& - (\beta -\gamma)(\sin k_z \sin k_x \sin k_y ) \sigma_x s_y , \label{eq:model_CdAs}
\end{align}
with $T=i \sigma_0 s_y K$ and 
$C_4=e^{i\frac{\pi}{4}(2+\sigma_z) s_z}$.
Here $M$, $t_{xy}$, $t_z$, $\eta$, $\beta$, and $\gamma$ are material-dependent real constants. 
If $t_z>(M-2t_{xy})>0$, this model has
a pair of Dirac points located at ${\bm k}=(0,0,\pm k_0)$, with $k_0>0$
defined by $M = t_z \cos k_0 +2 t_{xy}$.
We find $\nu_{\pm i}(0) = \pm 1$
and $|\mathcal{Q}|=1$.
Accordingly, the FL arises at the surface
parallel to the $k_z$ axis, and the Dirac points are protected by $C_4$.  

\begin{figure}[tbp]
\centering
 \includegraphics[width=8cm]{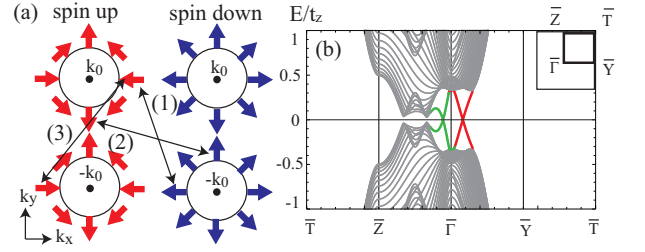}
 \caption{(Color online) 
(a) Orbital textures on the Fermi surfaces in the $k_x k_y$-plane with
 $k_z=\pm k_0$ in the $s_z=1$ (left) and $s_z=-1$ (right) sectors.
Arrors indicate the  direction of the orbit, $(\langle
 \sigma_x \rangle,\langle \sigma_y \rangle)$.
A Cooper pair between electrons with opposite spin may realize both of
a parallel orbit pair (1) and an anti-parallel one (2), depending on the
 momentum. On the other hand, a Cooper pair between electrons in the same
 spin state always have the anti-parallel orbit configuration (3).
(b) Energy spectra at the (100) face. $\mu/t_z = 0.5$, $M/t_z = 4$, $t_{xy}/t_z =2$, $\eta/t_z=1$, $\beta/t_z = 2$, $\gamma/t_z=1$,
 $\Delta_0/t_z=0.1$, $\Delta_0'/t_z = 0.01$, and $m_0/t_z = 0.2$. The distance
 between left ($x=0$) and right ($x=L$) surfaces is $L=50$. 
The red and green lines show
 the FL induced MFs.} 
\label{fig:Majorana}
\end{figure}

Near the Dirac points at ${\bm k}=(0,0,\pm k_0)$, the low-energy
Hamiltonian takes the form of the Dirac Hamiltonian,
$
H({\bm k})=\pm t_z k_0 (k_z\mp k_0)\sigma_zs_0
+\eta(k_x\sigma_x s_z-k_y\sigma_ys_0), 
$
which exhibits nontrivial {\it
orbit--momentum locking}. 
In Fig.~\ref{fig:Majorana}(a), we show orbital
textures in the $k_x k_y$ plane with $k_z=\pm k_0$ in each spin sector, where
the orientation of the orbit is tightly locked to the direction of
the momentum on the Fermi surface. 
Orbit--momentum locking critically affects the
possible pairing symmetry in the superconducting state. 
Indeed, we can show that constant $s$-wave pairing is
inconsistent with the orbital texture. 
First, in such a static pairing state, electrons forming the
Cooper pair have opposite momentum to each other, so they must belong to
different Dirac points.
Furthermore, 
as an $s$-wave pairing
is spin-singlet, it must be formed
between electrons in different spin sectors.  
However, for a Cooper pair between electrons in 
different Dirac points and different spin sectors, 
orbit--momentum locking requires a momentum-dependent orbital
structure in the Cooper pairing, as illustrated in
Fig.~\ref{fig:Majorana}(a). 
Therefore, even if the pairing interaction favors an $s$-wave
superconducting state, the pairing function cannot be constant, 
suggesting suppression of the critical temperature.

On the other hand, for a Cooper pair with parallel spins,
orbit--momentum locking is consistent with a constant
pairing function. 
Indeed, the orbital-singlet equal-spin pairing, 
$\Delta= \Delta_0
(c_{\uparrow,1} c_{\uparrow,2}-c_{\downarrow,2} c_{\downarrow,1}) 
+ i \Delta_0'(c_{\uparrow,1} c_{\uparrow,2}+c_{\downarrow,2} c_{\downarrow,1}) 
(\equiv\Delta_{\parallel}) $
is compatible
with the orbital texture in Fig.~\ref{fig:Majorana}(a).
Such an orbital-singlet Cooper pair is realized when the
effective pairing interaction is dominated by an attractive interorbital interaction
$
{\cal H}_{\rm int}=-2Vn_1n_2 
$, 
with $n_\sigma=\sum_{s=\uparrow, \downarrow}c_{s,\sigma}^{\dagger}({\bm
x})c_{s,\sigma}({\bm x})$ ($V>0$) \cite{Fu:2010,Nakosai:2012,Mizushima:2014,Brydon:2014}.
Although the actual pairing interaction is
material-dependent, the above results imply that doped Dirac semimetals favor the latter gap function. 
Because $\Delta_{\parallel}$ is $C_4$-symmetric
with $r=2$ and mirror-odd, i.e., $C_4 \Delta_{\parallel} C_4^t = - \Delta_{\parallel}$, it realizes
a symmetry-protected TSC with 
bulk point nodes and a surface MF quartet,  
as discussed previously.
In Fig.~\ref{fig:Majorana}(b), 
we illustrate the quartet of MFs in this phase 
by numerically calculating the surface energy spectrum of the BdG
Hamiltonian with Eq.~(\ref{eq:model_CdAs}) and $\Delta =
\Delta_{\parallel}$. 
Here we have also taken into account a
symmetry-lowering effect from $C_4$ to $C_2$ by phenomenologically adding $(m_0 \sin k_z ) \sigma_x s_x$ to Eq.~(\ref{eq:model_CdAs}). 
As expected, Fig.~\ref{fig:Majorana}(b) proves
the existence of MFs on the mirror-invariant line ($\bar{\Gamma}
\bar{Y}$) and the $k_z$ axis ($\bar{\Gamma}\bar{Z}$) and moreover shows a
gap in the $\bar{Z} \bar{\Gamma}$
direction due to the $C_4$ breaking term~\cite{footnote1}. The obtained MFs in the mirror invariant plane stay gapless even if the interaction effects are take into account~\cite{Gu:2014,Qi:2015}.

Finally, we discuss possible application of our theory to
superconductivity in Au$_2$Pb and Cd$_3$As$_2$.
For Au$_2$Pb, 
first-principle calculations show that 
the Fermi level of this material is inside the gap of the Dirac
points~\cite{Schoop:2014};
thus, 
no electron near the Dirac points contributes to the
Cooper pairs. 
Hence, no TSC as discussed above is expected in Au$_2$Pb. 
On the other hand, the analysis above is applicable to the recently
discovered superconductor Cd$_3$As$_2$.
In Cd$_3$As$_2$, under high pressure, the structural phase
transition occurs before the superconducting transition.
Together with the orbit--momentum locking discussed above,
the symmetry-lowering effect may stabilize the TSC
 phase by increasing the condensation energy,
as the point nodes in the TSC phase are gapped when $C_4$ is reduced to
$C_2$.
Therefore, it is likely that Cd$_3$As$_2$ realizes the TSC phase.
The mirror-odd gap function of the TSC is detectable via
anomalous Josephson effects~\cite{Fu:2010,Yamakage:2013} 
with carefully
fabricated junctions.  


The authors are grateful to Y. Yanase for discussions.
This work is supported in part by a Grant-in Aid for Scientific Research from MEXT of Japan, ``Topological Quantum Phenomena,'' Grant No. 22103005 and ``Topological Material Science'' (No. 15H05855) KAKENHI on innovation areas from MEXT. S.K. acknowledges support from JSPS (Grant No. 256466). M.S. is supported by Grant-Aid for scientific Research B (Grant No. 25287085) from JSPS.


\begin{thebibliography}{999}
\bibitem{footnote1}
The obtained TSC cannot be considered as a stack of lower dimensional topological phases. In this sense, it is a strong phase.
\end{thebibliography}
\end{document}